\documentclass[twocolumn,amssymb, nobibnotes, floatfix, aps, prb]{revtex4-2}
\usepackage[utf8]{inputenc}
\usepackage[english]{babel}
\usepackage{amssymb}
\usepackage{amsmath}
\usepackage{graphicx}
\usepackage{color}
\usepackage{physics}
\usepackage{mathtools}
\renewcommand{\vec}[1]{\boldsymbol{#1}}
\usepackage{hyperref}
\graphicspath{{Figs/}}

\begin{document}

\title{The emergence of one-dimensional channels in marginal-angle twisted bilayer graphene}
\author{Niels R. Walet$^1$}
\email{Niels.Walet@manchester.ac.uk}
\homepage{http://bit.ly/nielswalet}
\author{Francisco Guinea$^{1,2}$}
\email{Francisco.Guinea@imdea.org}
\affiliation{$^1$Department of Physics and Astronomy, University of Manchester, Manchester, M13 9PY, UK}
\affiliation{$^2$Imdea Nanoscience, Faraday 9, 28015 Madrid, Spain}

\date{\today}

\begin{abstract}
We generalize the continuum model for Moir\'e structures made from twisted graphene layers, in order to include lattice relaxation and the formation of channels at very small (marginal) twist angles. We show that a precise description of the electronic structure at such small angles can be achieved by i) calculating first the relaxed atomic structure, ii) projecting the interlayer electronic hopping parameters using a suitable basis of Bloch states, and iii) increasing the number of harmonics in the continuum approximation to interlayer hopping. The results show a complex structure of quasi one dimensional states when a finite bias is applied.
\end{abstract}

\maketitle


The study of the effect of twisting on layered van-der-Waals  materials has opened a new avenue to control correlations in such systems. The appearance of superconducting and insulating phases at the magic angle (in the order of $1.1^\circ$) has been well documented\cite{Cetal18a,Cetal18b}, but the study of what happens at much smaller angles, and thus longer wave-length Moir\'e patterns, is of equal interest. Such ``marginal" twist angles have recently been studied experimentally \cite{yoo_atomic_2019,xu_giant_2019}. The effect of applying electronic gates to a twisted bilayer has first been studied in Ref.~\cite{chung_optical_2015} (note that the phase diagram of these systems can be modified by an applied electric field \cite{GLGS17}). 

As is well understood, the lattice structure relaxes, since areas with the energetically favorable AB (midpoint) alignment grow, and those with AA (identical) alignment reduce in size. A network of narrow channels (also described as ``interface solitons") separating the distinct regions with AB  and BA alignment develops. It is well known from the study of stacking defects in aligned graphene layers that we find two (pairs of) chiral edge modes at such interfaces, a problem that has been well studied, both theoretically and experimentally
\cite{martin_topological_2008,jung_valley-hall_2011,zarenia_chiral_2011,%
vaezi_topological_2013,zhang_valley_2013,li_spontaneous_2014,SGGNG14,yeh_gating_2014,%
cosma_trigonal_2015,ju_topological_2015,pelc_topologically_2015,%
yin_direct_2016,lee_zero-line_2016,li_gate-controlled_2016,%
lane_ballistic_2018,Retal18,cheng_manipulation_2018,sun_valley_2019,%
yang_gapped_2019}. In twisted bilayer systems such states hybridize with the electronic spectrum in the AB aligned regions, but on applying an electric bias between the two layers, to which the edge states are not sensitive, one should be able to create a gap for the AB states, thus liberating the channel states.
The channel approach has also been invoked to describe the interacting phases of TBG  \cite{wu_coupled-wire_2019,liu_pseudo_2019,wu_ferromagnetism_2019}

There are only a small number of papers where this specific problem is studied in some detail. 
An early calculation, which does not include lattice relaxation can be found in Ref.~\cite{san-jose_helical_2013} (note that lattice relaxation is crucial in determining the formation of one dimensional channels). In a more recent paper, an elegant analytic calculation for electrons within  channels only is presented \cite{efimkin_helical_2018}, but this is limited to a single chiral pair, and the scattering at channel intersections is treated phenomenologically. There is also a development of a model where a field theoretical approach is applied to the many-body physics of the channel network, but not in a way that leads to specific predictions for the channel states \cite{wu_coupled-wire_2019}. Finally, in the work by Hou \emph{et al} \cite{hou_current_2019} a model of the channels is made using a tight-binding model in a single layer graphene with a position-dependent on-site potential; this approach lacks some of the chiral and layer-coupling aspects. All of these models have some limitations; this clearly leaves room for an in depth analysis of the edge states present in a realistic description of such systems. On the other hand, the relaxation of the atomic positions which leads to the formation of channels has been extensively studied \cite{NK17,GuineaWalet2019}, as well as the oscillations around the relaxed positions \cite{KS19,O19}. 

These systems can be accessed experimentally as well; the work of Yoo \emph{et al} \cite{yoo_atomic_2019} studies such systems to angles as small as $0.1$ degree--but without an electronic bias, which we shall see is crucial to separate the channel physics--and the work by Xu \emph{et al} \cite{xu_giant_2019} and Rickhaus \emph{et al} \cite{Retal18} looks in detail at the effect of interlayer bias, but the theoretical analysis of these results is somewhat limited.

\begin{figure}
\includegraphics[width=\columnwidth]{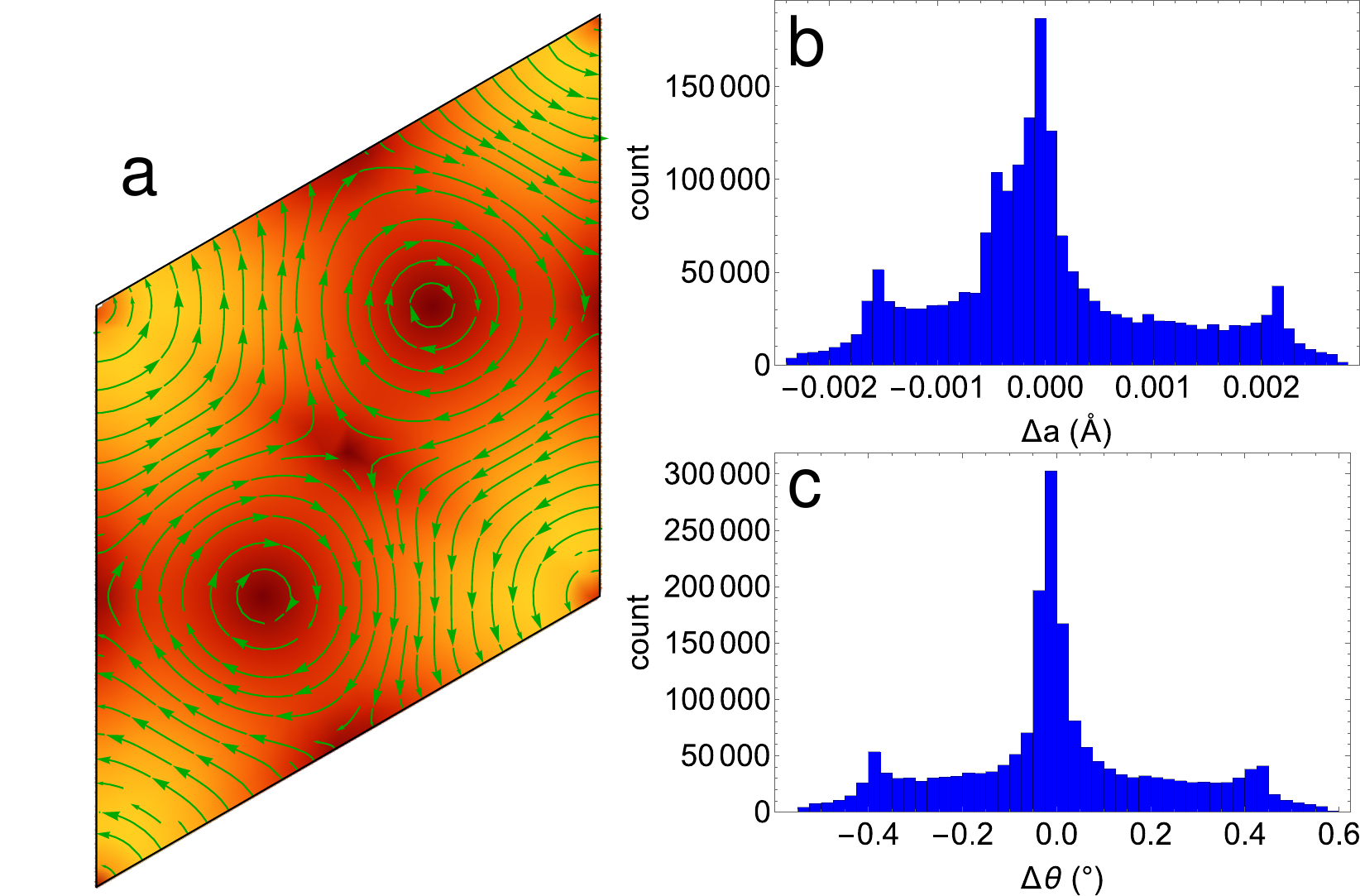}
\caption{A representation of the relaxation of a bilayer for a twist angle $\theta=0.221^\circ$.  This example is for the LCBOPI+KC potential, see Ref.~\cite{GuineaWalet2019}. a) Stream-line representation (in green) of the in-plane 
movement of atoms in the top layer on relaxation. A single rhombic unit cell is shown.  The background color shows the magnitude of movement: dark red for no movement, yellow for the largest movement. b) Bond-stretching of nearest neighbor bond from the equilibrium value $a=1.42\,\text{\AA}$. c) change of  bond angle from the equilibrium value $\theta=120^\circ$.\label{fig:relaxation}}
\end{figure}

The difficulty with approaching this problem more quantitatively is first of all that a direct tight-binding calculation is extremely time-consuming due to the large number of atoms in a unit cell ($268\,204$ for $\theta=0.221^\circ$), and secondly that the standard continuum model for a bilayer \cite{bistritzer_moire_2011} does not apply, see Ref.~\cite{GuineaWalet2019}. The problem is how to include  lattice relaxation in a tight-binding model, at the same time as taking into account the effect of the modification of the hopping due to the modified positions, and modified many-body screening effects. Fortunately, it seems to be a sensible approximation to assume that the in-layer electronic hopping parameters are unchanged, since the bond stretching is extremely small (less than a part in $10^4$), and the main deformation is torsional motion around the $AA$ aligned points--for a visual  representation see Fig.~\ref{fig:relaxation}.

Since the hopping parameters are essentially unchanged and the lattice has the same periodicity, the eigenfunctions for the layers are the same. As the modifications are small, any correction due to these can be included in first order perturbation theory. It remains to find the interlayer couplings: due to the periodic modulation of the Moir\'e pattern, we find that this allows a momentum transfer of a multiple of the superlattice reciprocal basis vectors. If one  makes an expansion in a sufficient number of reciprocal lattice vectors, it has been shown that we can reproduce the low-energy states of tight binding calculations with great accuracy \cite{GuineaWalet2019}.

In this note we shall apply this technique to investigate the behavior of marginal-angle twisted bilayer graphene. We shall start by looking at larger angles for a baseline comparison. We are then specifically interested in applying an electrical bias between the layers, which should enhance the effect of the chiral edge states along AB/BA interfaces \cite{san-jose_helical_2013,efimkin_helical_2018,wu_coupled-wire_2019}. According to results quoted in Ref.~\cite{efimkin_helical_2018}, a bias of up to $250\,\text{meV}$ is reachable, even though values of the order of $100\,\text{meV}$ seem more reasonable. Related work, although not focused on the one dimensional network of channel states, can be found in Refs.~\cite{CFZK19,FCZMK19}.

\section{Results}

\begin{figure}[tbh]
\includegraphics[width=0.8\columnwidth]{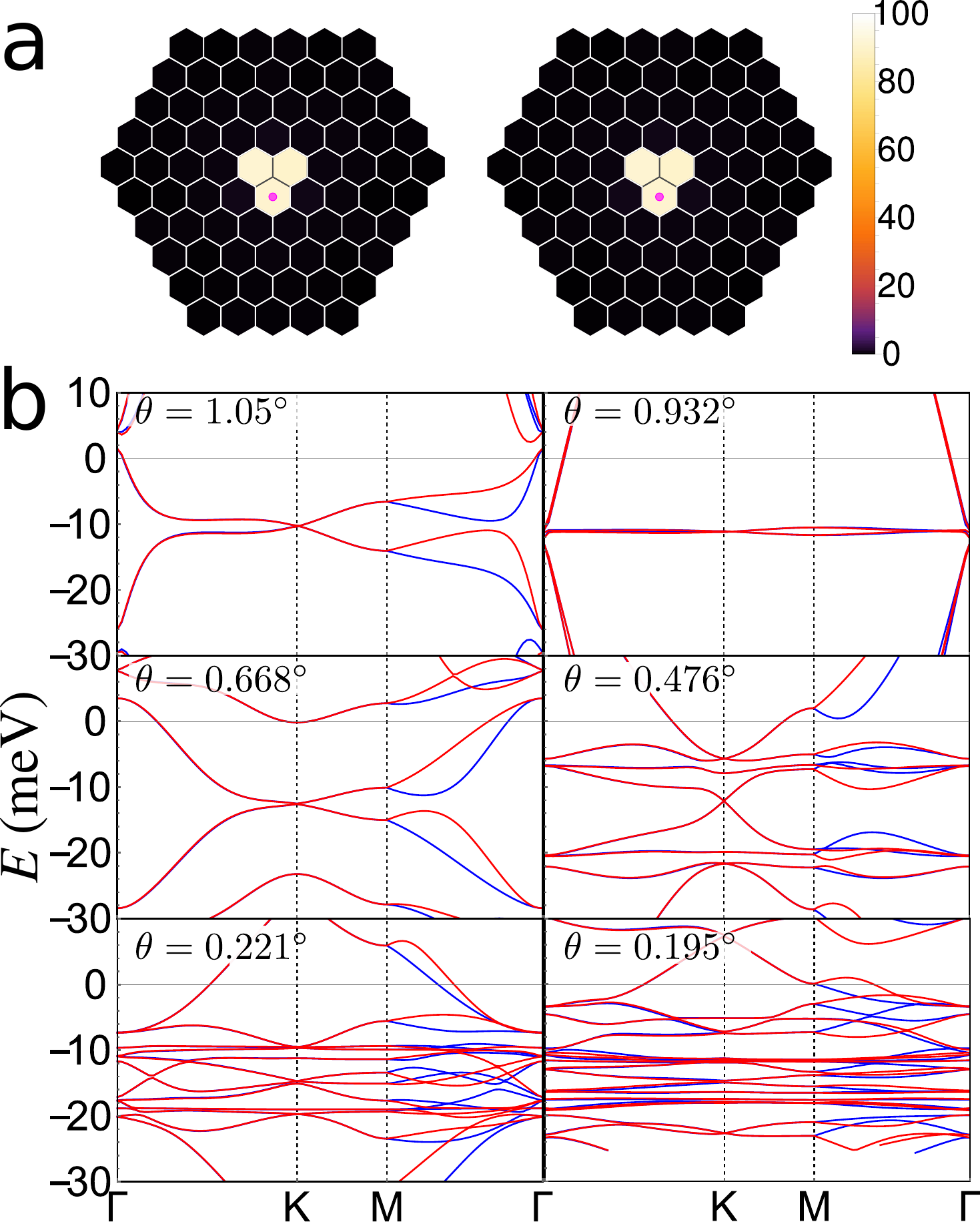}
\caption{a) The magnitude of the projected couplings for the continuum model  for  flat undeformed layers: $AA$ couplings on the left and $AB$ to the right. Each hexagon denotes a single reciprocal lattice vector; the one with the pink dot is $(0,0)$. b) Bands for the continuum projected model  for flat undeformed layers. Red and blue label the two valleys.}\label{fig:init}
\end{figure}

We study results for a pair of flat (fixed distance $d=3.46\,\text{\AA}$) graphene layers, which we have relaxed under influence of the ``LKC" (LCBOPI+KC) potential model discussed in detail in ref.~\cite{GuineaWalet2019}. Since we are mainly interested in the channel states in this paper, we shall not use the complex many-body screening discussed in that reference, but we shall only study the exponential Koster-Slater hopping of the form 
\begin{equation}
t(r_{12})=0.52 \exp(-2.2 [r_{12}-3.46])\,\text{eV}.
\end{equation}
Note that this has been enhanced by a factor of $1.3$ relative to our previous work, in order that the first magic angle occurs at $1.05^\circ$ for the flat LKC relaxed lattice.
As always, we start with the case of 
rigid graphene sheets, without any relaxation. 
We do not perform tight-binding calculation, but rather construct a continuum model from the tight-binding Hamiltonian using a method recently developed by us \cite{GuineaWalet2019}. This generalizes the now ubiquitous continuum model of twisted bilayer graphene \cite{bistritzer_moire_2011,LDSPCN12,SJGG12} to include more than a single triangle of couplings--these couplings are due to the momentum transferred by the misalignment of the two layers. We have shown in our previous work that this projection can be extremely effective in reproducing the low-energy spectrum of tight-binding calculations, but that for more complicated interlayer hopping parameters the number of terms in the expansion of the interaction (the number of harmonics, loosely "the number of superlattice vectors", see Fig.~\ref{fig:init}a ) that needs to be included gets rather large. For that reason we shall only employ the exponential ``Koster-Slater" form studied in our previous paper; we have used a scaled strength so that the magic angle is at $1.05^\circ$.

First we look at a baseline, the perfect Bistritzer-MacDonald  projection (with only three harmonics) for an undeformed lattice--this projection is independent of lattice size, as we have also checked numerically, see Fig.~\ref{fig:init}a) for a typical example. Note that our version of this model uses two additional features: a full in-lattice dispersion, rather than a linear expansion, and a first order momentum dependence of the interlayer coupling parameter--both of these are required to reproduce the shift of the Fermi energy away from zero seen in tight binding calculations.

As we can see in Fig.~\ref{fig:init}b), [see also Fig.~\ref{fig:band1i2}] as we decrease the angle we find an ever denser set of states near the Fermi energy, which lies in the order of $10\,\mathrm{meV}$ below zero due to electron-hole symmetry breaking. For the smallest angle used this gives a complex tangle of flat bands within $10\,\mathrm{meV}$. 

\begin{figure*}
\includegraphics[width=0.8\textwidth]{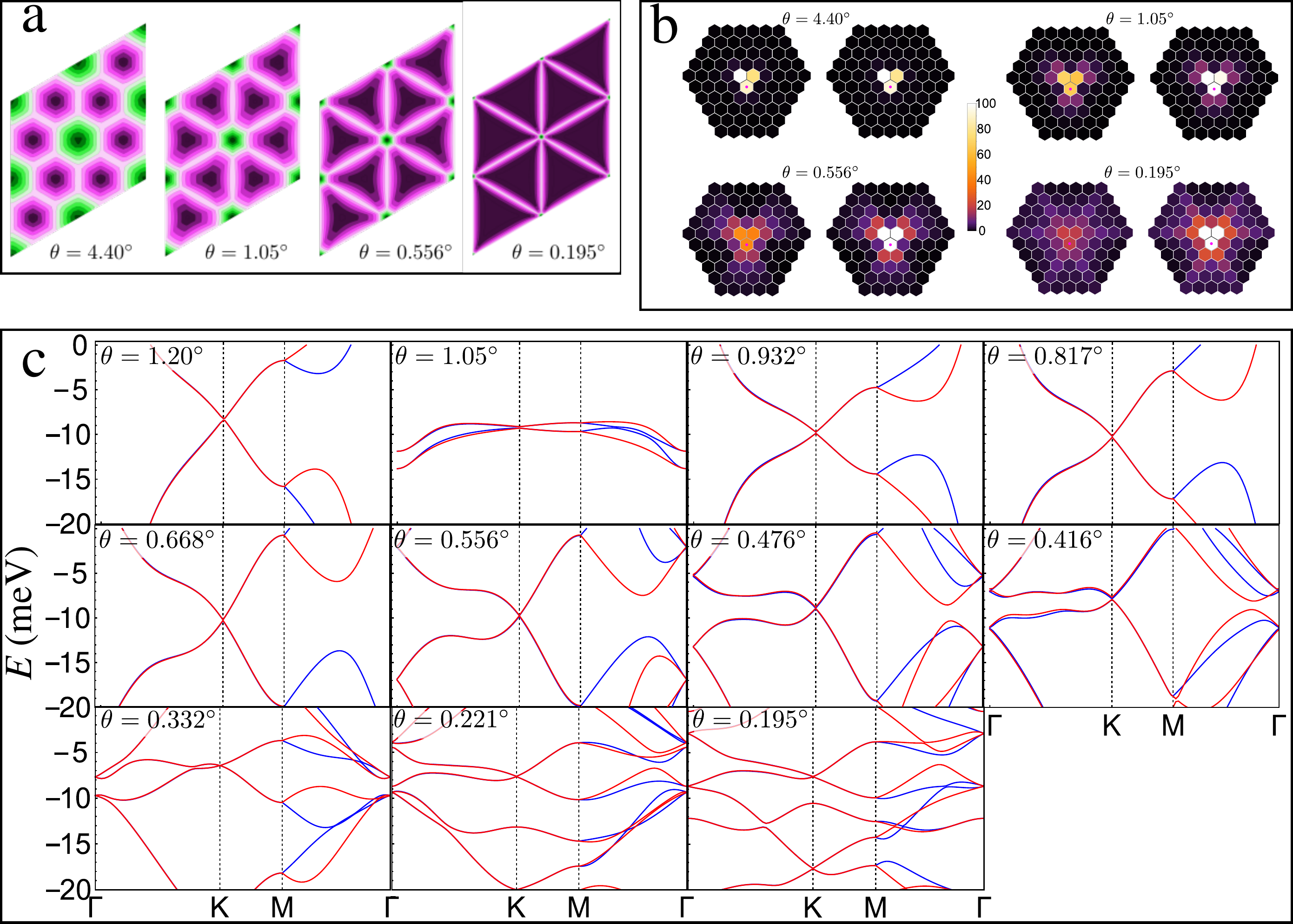}
\caption{a) The alignment measure from Ref.~\cite{GuineaWalet2019} for a flat relaxed bilayer. The green areas are $AA$ aligned; the purple ones $AB$. White means equal alignment, and will show the channels. These plots are  not to scale: the physical dimensions of these figures increase as we go to smaller angles, and can be gauged by the size of the green areas, which is roughly similar in all cases.
b) The magnitude of the projected couplings for the continuum model for a flat relaxed bilayer: $AA$ couplings on the left and $AB$ to the right. The hexagon with the pink dot is for a momentum transfer $(0,0)$. The next hexagons correspond to a transfer of one unit of superlattice momentum $G=2\pi/L$.
c) Bands for the continuum projected models  for a flat relaxed bilayer. Red and blue label the two valleys.
For more detail, see Figs.~\ref{fig:relaxed2},\ref{fig:hex2},\ref{fig:band1}.
\label{fig:relax}}
\end{figure*}

The situation changes rather drastically for a flat relaxed bilayer. 
The relaxation leads to an enhancement of regions with $AB$ and $BA$ alignment. If we plot this using the alignment measure developed in our previous work \cite{GuineaWalet2019}, we see that we evolve from an alignment that for the largest angle considered has roughly equal areas with $AA$ and $AB$ alignment to a situation with dominant $AB$ alignment, where we are left with very narrow channels connecting tiny $AA$ aligned regions  for  angles of the order of $0.2^\circ$, see Fig.~\ref{fig:relax}a, or for a more complete set see Fig.~\ref{fig:relaxed2}.

This means that the continuum couplings, as shown in Fig.~\ref{fig:relax}b and \ref{fig:hex2}, now do change with the twist-angle: the $AA$ coupling go down rapidly with angle, and the range of the $AB$ couplings increases with angle, even though still dominated by the three central harmonics of roughly constant strength independent of twist angle.
The increase in the area with $AB$ alignment (see Fig.~\ref{fig:relax}a) is largely responsible for this. The change in the coupling matrix elements also leads to very different spectra--surprisingly enough, even for the flat bands at $\theta=1.05^\circ$ this effect is already  pronounced, but the effect seems to increase as we go to larger lattices. [The results for large angles are likely to be incorrect due to the neglect of inter-valley coupling, which is absent in the continuum model, but non-negligible for such systems.] 

The bands are modified due to the relaxation, but near the first magic angle they seem only marginally different from the bands for the unrelaxed lattice, see Fig.~\ref{fig:relax}c, 
but for smaller angles the spectrum seems less dense than in Fig.~\ref{fig:init}b.

\begin{figure}
\includegraphics[width=0.7\columnwidth]{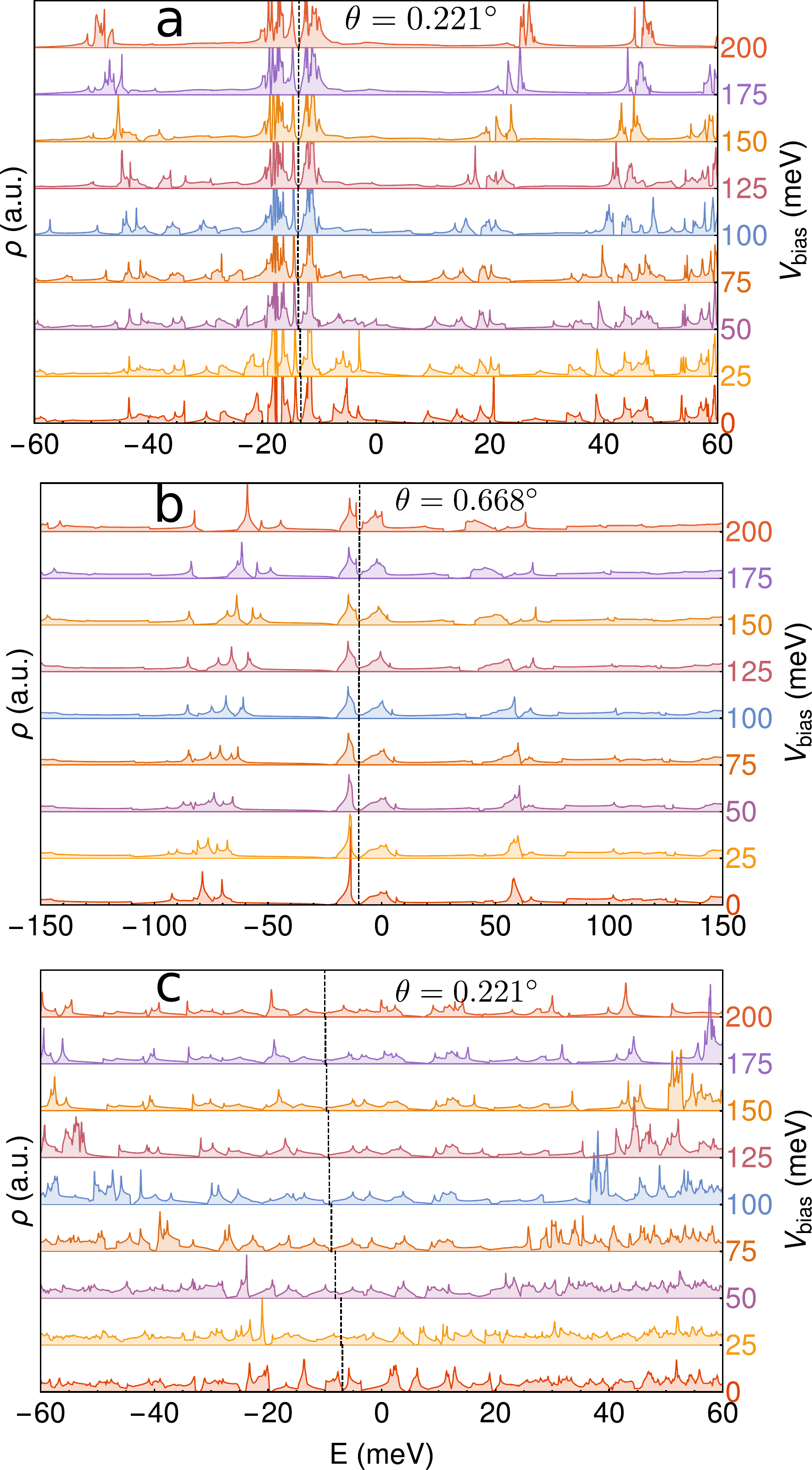}
\caption{a) Density of states for the continuum projected models with a bias ranging $0$ to $200\,\text{meV}$ for flat undeformed layers at twist angle $\theta=0.221^\circ$. b) and c) Similar for for a flat relaxed bilayer.  The black dashed lines indicate the Fermi level.}\label{fig:DOSbias}
\end{figure}

If we have channel states, which are hybridized topological states between the two layers, we expect that these are not sensitive to an applied interlayer bias by gating the top and bottom layer. We would expect that those states located largely in $AB$ regions are very sensitive to such a bias, and we can thus untangle the two parts of the spectrum. For layers without relaxation, surprisingly little sensitivity occurs to such a bias. The channels in this case form a hexagonal network (with rather broad channels), and there seems to be an indication that such a network has zero-energy states, with a strong set of van Hove singularities near the Fermi surface, that are hybridised, but no separate $AB$ states, as can be seen in Fig.~\ref{fig:DOSbias}a. What we notice there is the the spectrum hardly budges as we change the bias from $0$ to $200\,\mathrm{meV}$. There is a small amount of spectral density near $E_F$ that seems to be moved, but most of the spectrum is invariant. A larger selection of relevant results are shown in Fig.~\ref{fig:DOSbias1a}.


That raises the question what happens for the fully relaxed planar layers. We show two typical cases in Fig.~\ref{fig:DOSbias}b and c. For  angles larger than about $0.5^\circ$, we see little sensitivity to an applied bias. For smaller angles we see an interesting pattern emerging: whereas we have rather busy and flat structure for $V=0$, we find that an applied bias gradually moves a subset of states, most likely the $AB$ aligned ones, to higher and lower energies, liberating a set of states that are not  very sensitive to the applied bias. There is a clear tipping point, as can be seen in Fig.~\ref{fig:DOSbias2a}, around $0.5^\circ$.

\begin{figure}
\includegraphics[width=\columnwidth]{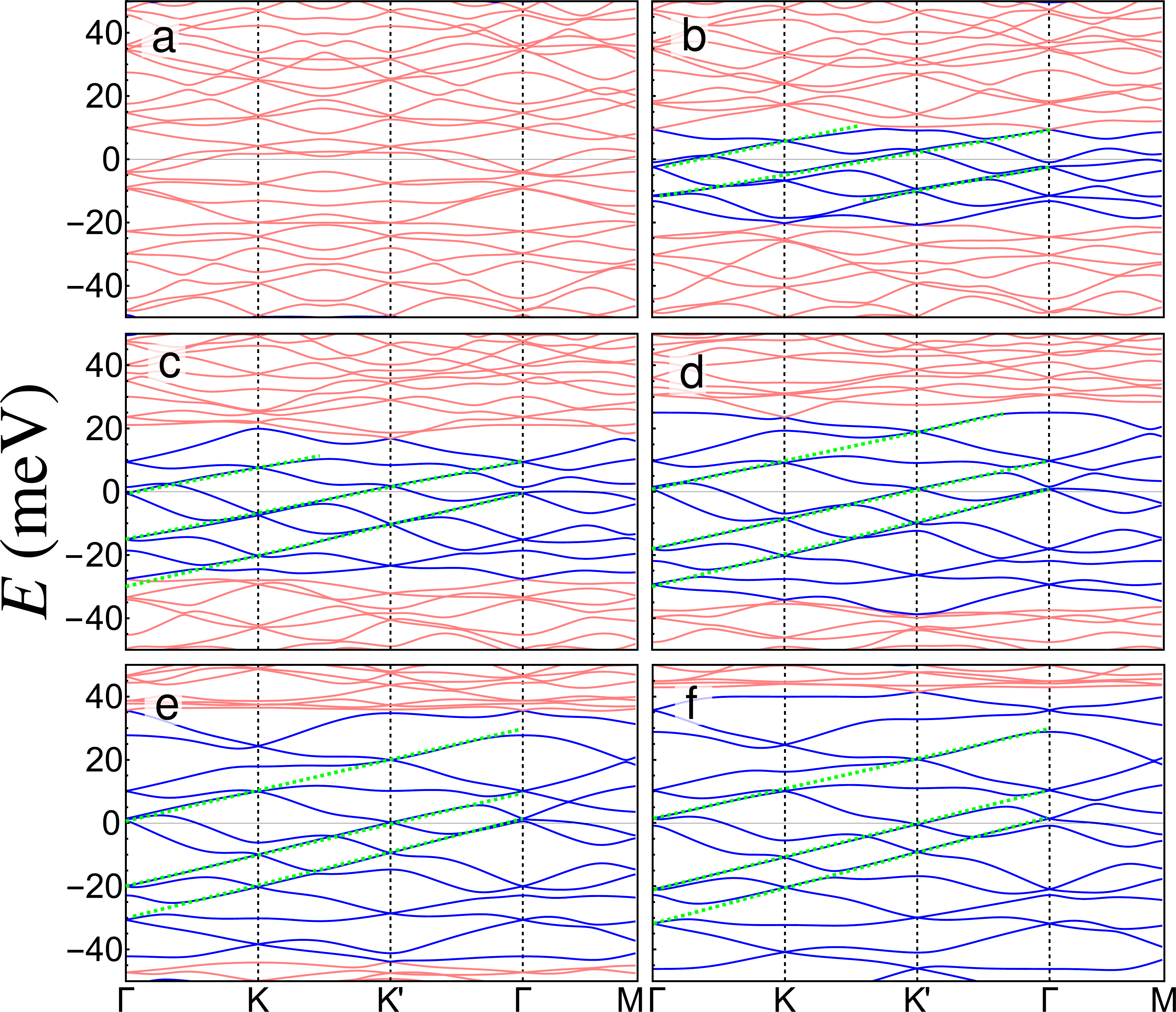}
\caption{Bands for the continuum projected models with a bias (a: $0$, b: $25$, c: $50$, d: $75$, e: $100$ and f: $125\,\text{meV}$) for a flat relaxed bilayer at a twist angle of $0.221^\circ$. We show results for a single valley; the solid blue lines label states we identify as channel states, and the light red color labels likely non-channel states.  The green dotted lines indicate a potential identification of up-sloping bands.}\label{fig:biasflat}
\end{figure}

This becomes more explicit if we look at Fig.~\ref{fig:biasflat}. There we  have attempted to color the states according to whether they are in a region where states are moving with bias or not [there is some uncertainty in this process at the edges]. We have only shown one valley; the other valley looks very much the same, apart from a trivial mirroring along the midpoint $M$ between $K$ and $K'$, since we now have the opposite chirality. See also the video Fig.~\ref{vid:interest} for a dynamical picture how the channels develop with increasing bias. A few additional examples can be seen in Fig.~\ref{fig:bias1a}.
What we see is that there is a stable network of states developing, with interesting features.

These features are most easily seen if we compare this to results from the Efimkin-McDonald (EM) model \cite{efimkin_helical_2018}. In that model, which considers a single chiral edge state, we see that we have a large number of parallel straight upward sloping bands, intersected by undulating slowly descending bands, see Fig.~\ref{fig:EmcD}. The parameter $\alpha$ of that model (where $P=\cos^2\alpha$ is the fraction of flux transmitted in the forward direction at the $AA$ junctions) is argued to be large, so that the intersection between undulating and straight bands is quite visible. Our results seem to suggest that this is probably the case, but differ in important aspects.
First of all, we only see one pair of up-sloping lines--maybe there is a little indication of a third. Also, the crossings between the undulating and straight lines occur at the $K$, $K'$ and near the $M$ point--the latter is completely absent in the EM model. This suggests that we can create two groups of straight and undulating lines, with some couplings. This probably means that each of these are build on one of the chiral bands expected at the edge. It may be the case that the interaction between these two is less important at low energy, or that the finite width of the channels limits the validity of any 1D model to the lowest energy, and that higher states are quasi-1D hybrids.

\begin{figure}
\includegraphics[width=\columnwidth]{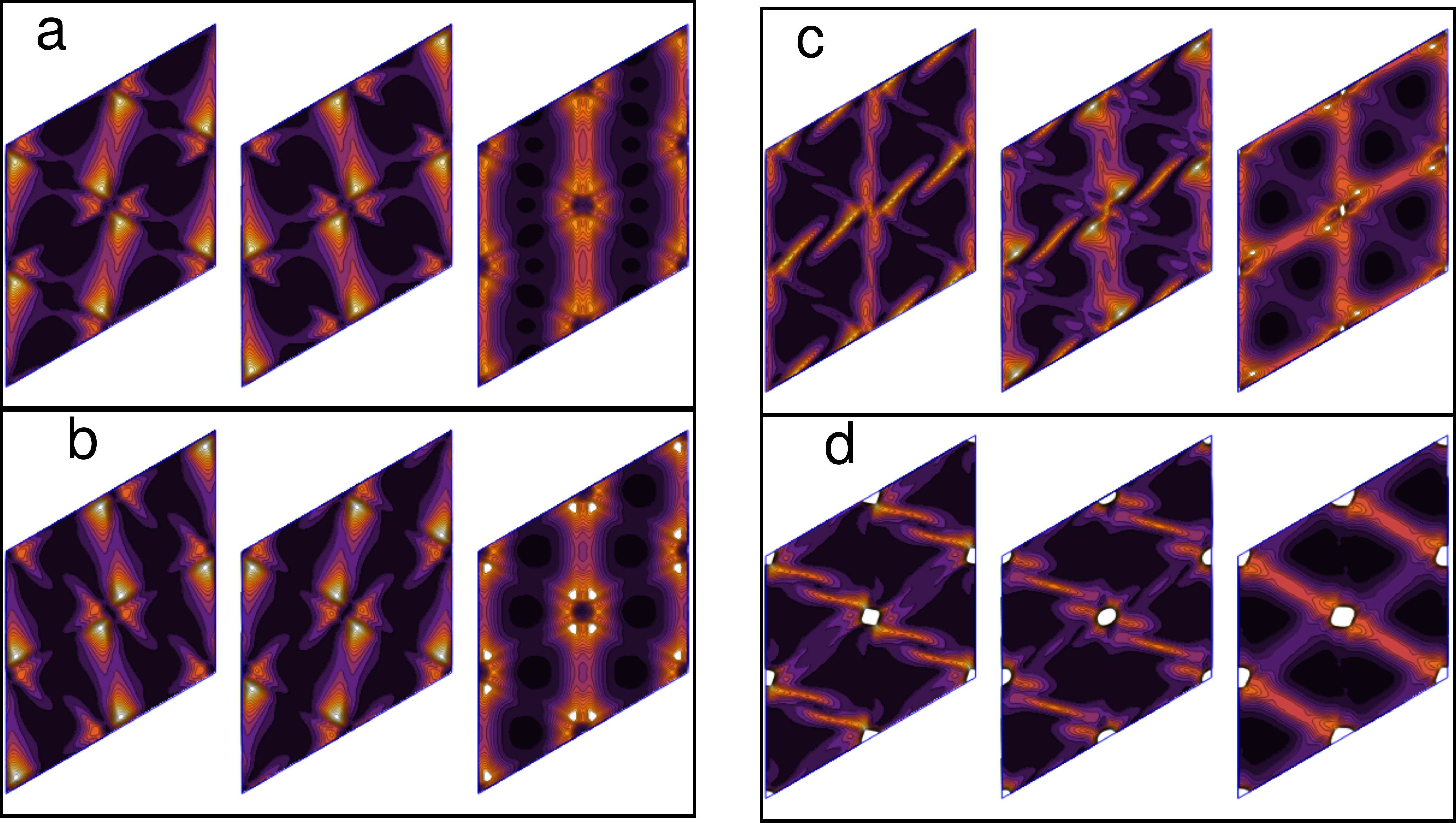}
\caption{The probability density for states near zero energy at the $M$ point for  an angle of $0.221^\circ$. a-b:  \emph{no} applied bias; c-d: bias of $100\,\text{meV}$.
The energy eigenvalues are a) $E=-10.2$ b) $E=-3.95$ c) $E=-16.09$ and d) $E=-14.81\text{meV}$. In each panel left is the top-layer wave function, middle the bottom layer one, and the right panel is the total density.\label{fig:pall}}
\end{figure}

A study of the wave functions should be able to shed some light on these problems. We first look at the case without a bias, and investigate the wave functions at the $M$ point of the Brillouin zone. \footnote{In order to be able to overlay the probabilities, which are each defined on different lattice points, we have interpolated separately the wave functions for each of the four sublattices (top $A$ and top $B$, bottom $A$ and $B$) before adding the relevant wave functions in quadrature.} As we see in Fig.~\ref{fig:pall}a-b and \ref{fig:palla}a-f, in this case  the wave functions show some sensitivity to the channels, but are not located mainly in these features: they are spread over a relatively large area of the real-space lattice. In  half the cases we see strong peaks in the small $AA$ aligned areas.

The positions of the wave functions change as we apply a bias. In all cases shown in Fig.~\ref{fig:pall}c-d  and \ref{fig:palla}g-l we see a strong concentration on the channels, with beats in the wave function between the top and bottom layers, showing the effect of standing waves in the channels. These beats also have a chiral pattern: this is linked to the valley under consideration, and the chirality is opposite in the other valley. In almost all cases we find a substantial fraction of the wave function located in the $AA$ region. This is due to the fact that there is substantially more binding in the small $AA$ region than along the rest of the channel: if we zoom in on individual channels, we also notice that the midpoint of each channel has the lowest probability density along the channel. This suggests a model with a strong coupling between the channels in both layers, with an attractive potential in the $AA$ region, and a slowly varying potential along the channel, with its maximum at the midpoint. Neither of these is present in the model by Efimkin and MacDonald \cite{efimkin_helical_2018}, and it is unlikely that this can and was described correctly in the single layer model of Ref.~\cite{hou_current_2019}; a single layer model can definitely not describe the interlayer beats and their chirality.

We look in a bit more detail at cases b) and c), which should correspond to the two almost straight lines in the spectrum (in the case b), this is near the gap in this line, so may be less convincing). We see that these look superficially rather similar, since the wave function vanishes in one of the three channel directions. There are some differences between these two cases, and b) is largely a single channel case. This may be due to the nearby avoided crossing with the states shown in a); however we see no sign of the vertical channel that is so prominent in that case.

Our model is unique, in the fact that it is simple, and seems to give a description that is in between the 
Efimkin and MacDonald one \cite{efimkin_helical_2018}, that has no notion of a gap, but concentrated on  describing one-dimensional channels connected through three-fold junctions, and the Zhang-McDonald-Mele model \cite{zhang_valley_2013}, which is very powerful for describing the chiral edge states, but where it is difficult to include the $AA$ scattering centers at the intersection of the channels [See supplementary material]. Actually, the ZMM model can also be seen as a generalisation of the model in Ref.~\cite{hou_current_2019}.

\section{Conclusions}

We have generalized the continuum model of the electronic bands of twisted graphene bilayers to include the effect of lattice relaxation at small twist angles. By including a suitable number of interlayer harmonics, the continuum model can be applied to small twist angles with no significant increase in computational cost. Calculations with accuracy comparable to that obtained for twist angles $\theta \gtrsim 1^\circ$ (see\cite{bistritzer_moire_2011}) can be achieved for angle $\theta \approx 0.1^\circ$, where the number of atoms in the Moir|'e unit cell is $\sim 10^6$.

We have performed an exhaustive study of the low energy part of the spectrum in the presence of an interlayer electric bias. We have focused on the existence and nature of quasi one dimensional states in small angle twisted bilayers in a perpendicular electric field. 
We find that the low energy bands, and the density of states tend to a bias independent limit, consistently with a situation where only channel states exist. On the other hand, our results show differences with more simplified models where only one set of modes per channel are studied\cite{efimkin_helical_2018}. The three band periodicity obtained in this calculation cannot be observed in our results, suggesting that the parameters such as the velocity of the electrons in the channels, and the scattering at the channel crossings, are energy dependent. It is also worth mentioning that our results show a small number of bands with almost linear dispersion, which describe electrons which move at uniform velocity throughout the lattice, experiencing very little Bragg scattering.

We have largely concentrated on one (simple) model of bilayer graphene. As we have argued in our previous work \cite{GuineaWalet2019}, this is likely to be too simple. Therefore, we have also performed calculations for another model of the relaxations, and for an alternative interlayer hopping. In all these cases we see a similar behavior, even though some details are different. We thus conclude that the occurrence of channel states is a robust result of relaxation in bilayer graphene. The details depend both on the theoretical model, which is also linked to the detail of the experimental set-up: mounting the graphene bilayer on or even encapsulating it with hBN will give a very different behavior than a free-standing bilayer. All of these details deserve consideration.

\section*{Methods}
The technique used to perform the calculations shown in this model is the modified version of the Bistritzer-McDonald method \cite{GuineaWalet2019}. In each case we have include 75 $\vec{G}$ vectors, and used a basis of 271 states to perform diagonalisation, unless stated otherwise.
\acknowledgments
NRW acknowledges support by the UK STFC under grant ST/P004423/1.
The work of FG is supported by funding from the European Commission under the Graphene Flagship, contract CNECTICT-604391.
\section*{Author Contributions}
Both Authors contributed equally to the design of the research and the writing of the paper; NRW performed most calculations.
\section*{Competing Interests statement}
The authors declare no competing interests.
\bibliography{GrapheneDomainWalls}

\clearpage
\appendix
\renewcommand\thefigure{S\arabic{figure}}    
\renewcommand\theequation{S\arabic{equation}}    
\setcounter{figure}{0}   
\setcounter{equation}{0}   

\section{Models of chiral bands}
\subsection{Efimkin-McDonald model}
\begin{figure}[htb]
\includegraphics[width=\columnwidth]{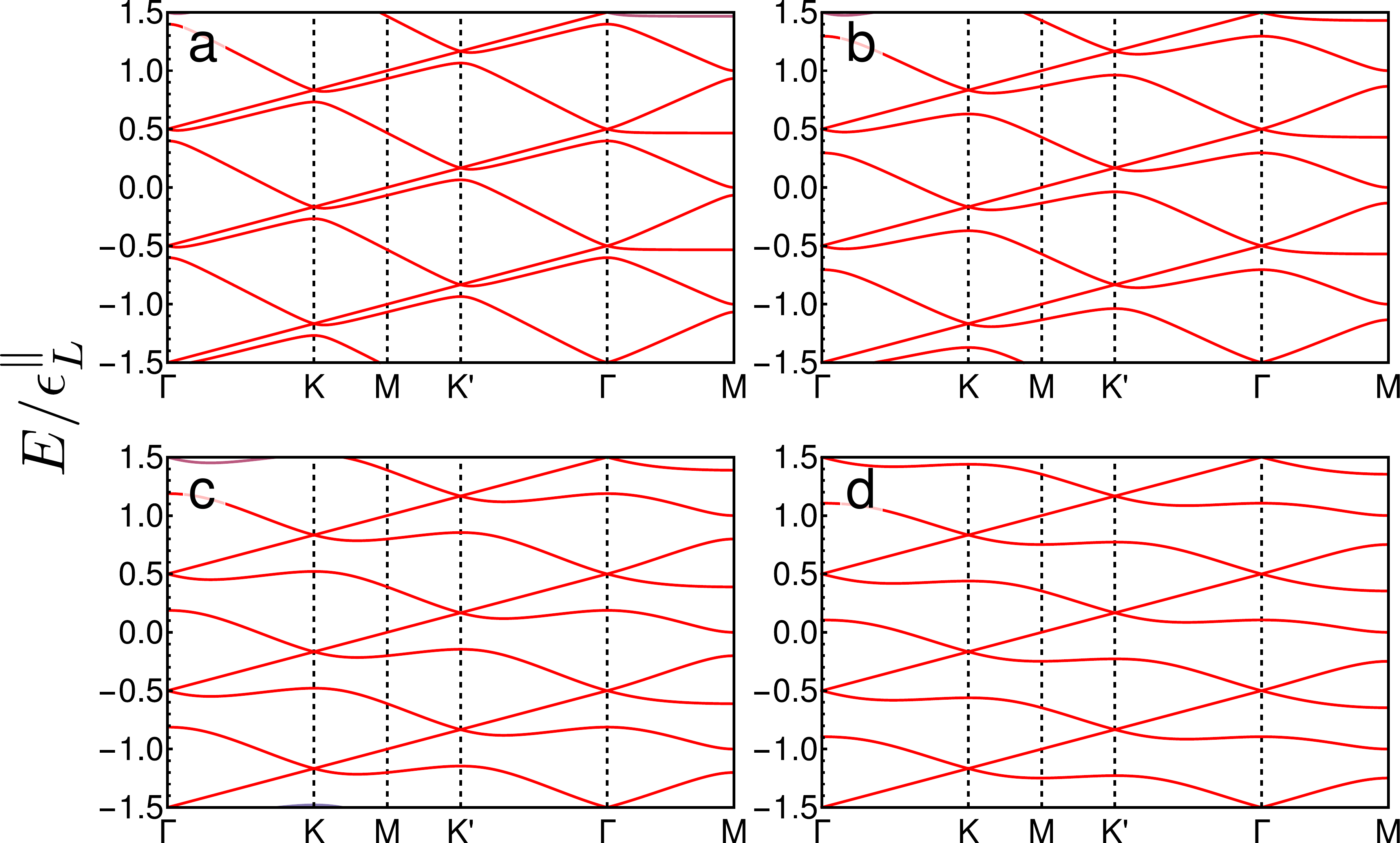}
\caption{Results of a calculation for the Efimkin-McDonald model \cite{efimkin_helical_2018} for angles $\alpha=0.3$ (a), $\alpha=0.6$ (b), $\alpha=0.9$ (c) and $\alpha=1.1$ (d). Here small $\alpha$ is largely forward scattering, and large $\alpha$ is mainly sideways scattering. The quantity $\epsilon^\parallel_L$ i expresses the natural energy scale of the model.}\label{fig:EmcD}
\end{figure}

In the Efimkin-McDonald model \cite{efimkin_helical_2018}, the spectrum is (apart from a constant shift in energy) determined by the ballistic propagation of electrons along a channel, with scattering in all directions at the $AA$ joints. Apart from some phases that only lead to a constant sift of the zero of energy, this is described by single angle $\alpha$ that gives the ratio of sideways to forward scattering: $P=\cos^2\alpha$ is the fraction of flux transmitted in the forward direction at the $AA$ junctions.
We show results along the same high-symmetry lines as in the main text in Fig.~\ref{fig:EmcD}. We note that the large angle results look more like the results in the main text. The key difference is the crossings occur only at the $K$ and $K'$ points. This is linked to the fact that the Efimkin-McDonald model only describes a single helical state.

\subsection{Zhang-McDonald-Mele model}
The ZMM model \cite{zhang_valley_2013} was instrumental in showing the fact that we have two chiral states at the interface between $AB$ and $BA$ alignment. It is essentially a continuum model of aligned bilayer graphene,
\begin{equation}
    H=\nu  \tau _0 q_x \sigma _x+\tau _0 q_y \sigma _y+\frac{1}{2} \gamma  \left(\sigma _x \tau _x-\mu \sigma _y \tau _y\right)+\Delta \sigma _0 \tau _z\,,\label{eq:ZMM}
\end{equation}
expanded about the $K$  ($K'$) valley for $\mu=1$ ($\mu=-1$). The value of $\nu$ labels the alignment, and is $\pm1$; we can of course allow jumps from one region to another. This model can be solved, numerically or analytically, for the gap along a straight edge separating two semi-infinite domains with $\nu=+1$ and $\nu=-1$, and shows a pair of valley-locked chiral states, with opposite chirality for each valley.

\begin{figure}[tbh]
\includegraphics[width=0.8 \columnwidth]{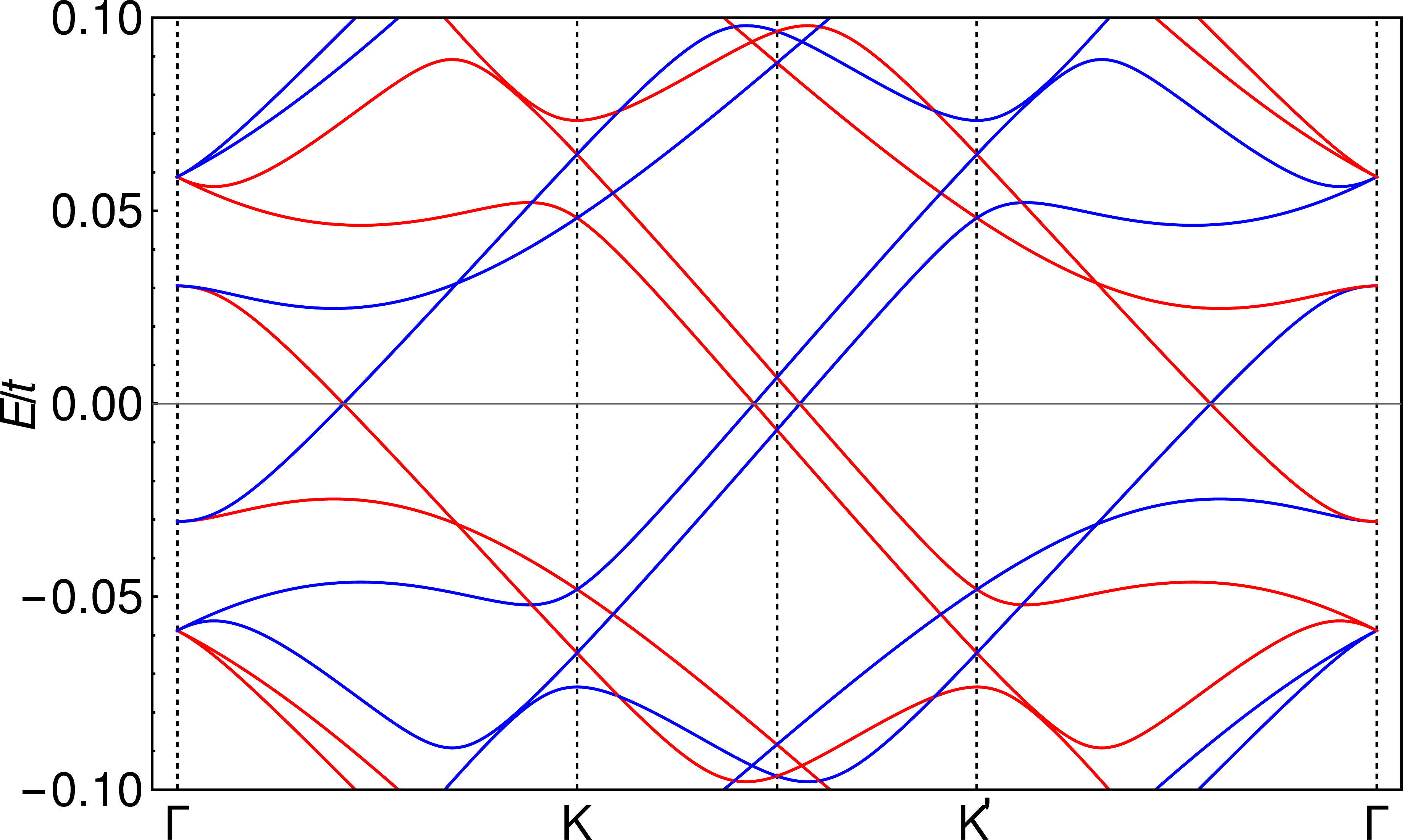}
\caption{The ZMM model for a $20$ lattice units supercell,
$\Delta/t=0.05$, $\gamma/t=0.1$ for zero width channels.}\label{fig:ZMM}
\end{figure}

There are multiple ways to turn Eq.~(\ref{eq:ZMM}) into an effective tight-binding model (or finite difference discretisation), which can then be used on more complicated lattices. To preserve the symmetry of our channels, we choose to work on an hexagonal lattice; the most elegant way to do so, set out below, gives rise to two uncoupled realisations of the channels, which differ slightly but converge in the continuum limit.
The reason is that on each site we have a 
$4\times 4$ model that describes both $A$ and $B$ on both layers, with the symmetry broken by the alignment term proportional to $\gamma$, and thus either the $1$ and $4$ or the $2$ and $3$ components couple. The onsite Hamiltonian is
\[H^\text{on-site}=\begin{pmatrix}
\Delta &0& 0&\frac{1}{2} \gamma  (1+\mu)\\
0 & \Delta &\frac{1}{2} \gamma  (1-\mu)& 0\\
0 & \frac{1}{2} \gamma  (1-\mu) & -\Delta & 0 \\
\frac{1}{2} \gamma  (1+\mu) & 0 & 0&-\Delta \\
\end{pmatrix},\]
where $\mu$ is site dependent, but only takes the values $\pm1$ (zero width channels).  If we use a standard labelling where $ij$ labels sites on a triangular lattice, and $k=1,2$ labels a shift between $A$ and $B$ lattice points, we find a hopping Hamiltonian ($\bar 1=2$ and $\bar 2=1$)
 \begin{align*}
H^{\text{hopping}}_{{ijk,i'j'k'}}=&
\delta_{k\,\bar{k'}}
(\delta_{ii'}\delta_{jj'}+\delta_{ii'}\delta_{j-j',k-k'}+\delta_{jj'}\delta_{i-i',k-k'})\\
&\times 
\begin{pmatrix}
0 &2t/3& 0&0\\
2t/3 & 0&0& 0\\
0 & 0& 0 & 2t/3 \\
0 & 0 & 2t/3&0 \\
\end{pmatrix},
\end{align*}
where $\vec{r}_{ijk}=i\vec{a}_1+j\vec{a_2}+(k-1)\vec{\delta}_1$.
This can now be applied to a rhombic supercell of a sensible size, and a reasonable choice of parameters (all expressed in terms of $t=1$). As we can see in Fig.~\ref{fig:ZMM}, we get states located in either valley (It is a minor nuisance that it is not trivial to separate the valley states, but looking at all the crossings, we see that the states for each valley do not mix). We see exactly the same number of up and down-sloping lines. The lack of scattering centres probably means that we lack the standing waves that are so typical of the realistic situation, see  main text. If we broaden the channels, gaps develop in the spectrum, again rather similar to what we have seen for the complete model.

The two models shown above are thus in some sense complimentary approaches to the more complex results of our continuum model. The Zhang \emph{et al} model seems of a similar computational complexity as ours once we discretise, but we find it hard to capture all the details of the full calculation in this model, apart from the finite number of states with a rising energy. The Efimkin-McDonald model is simple, but lacks the concept of the gapped $AB$ aligned states, and only describes a single chiral state.

\begin{widetext}
\newpage

\section{Supplementary figures}

\begin{figure*}[h]
\includegraphics[width=\textwidth]{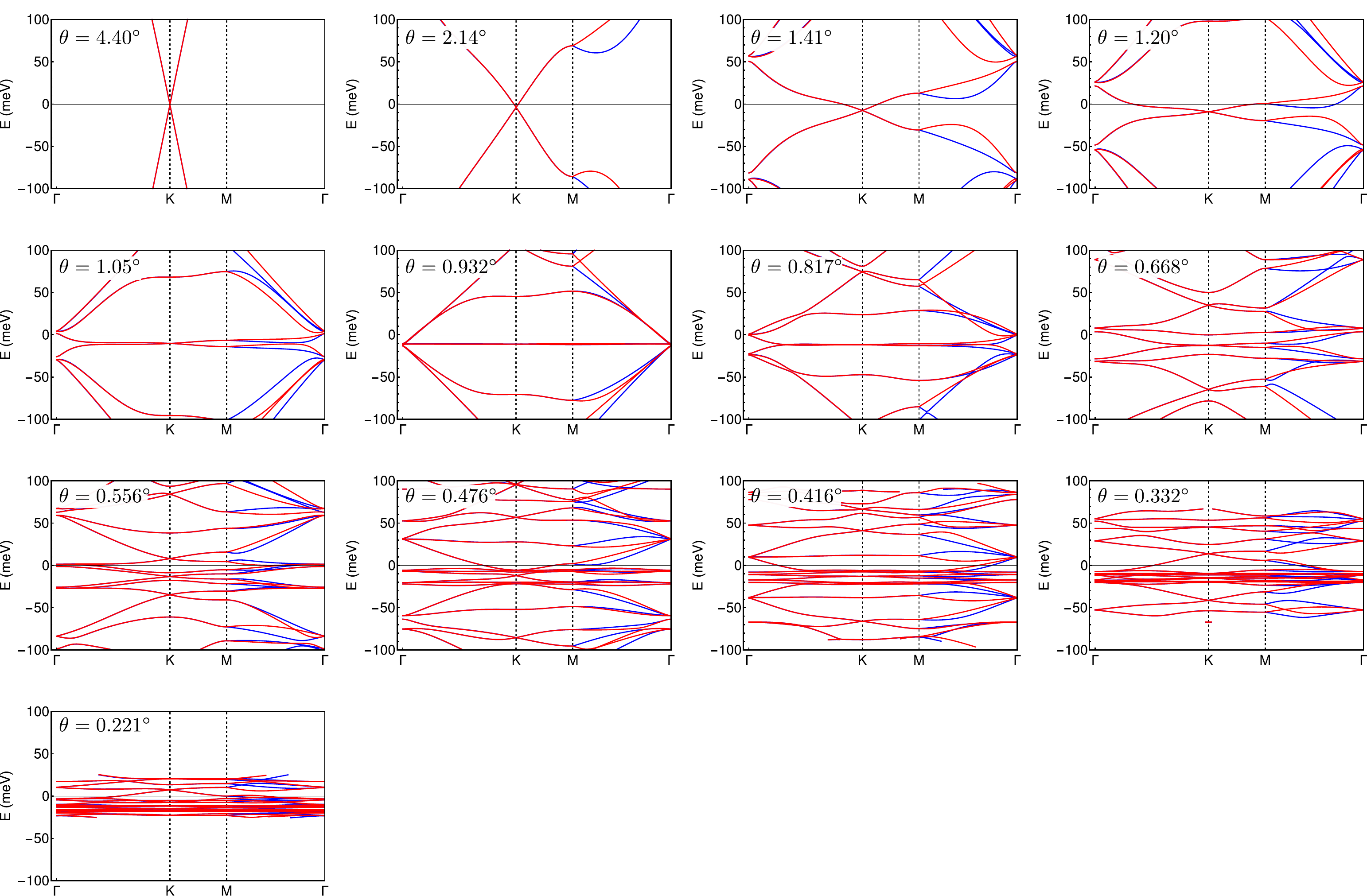}
\caption{Low-energy bands for the continuum projected model  for flat undeformed layers. We show both valleys, coloured red and blue--only a limited number of eigenvalues are shown.}\label{fig:band1i2}
\end{figure*}

\begin{figure*}
\includegraphics[width=\textwidth]{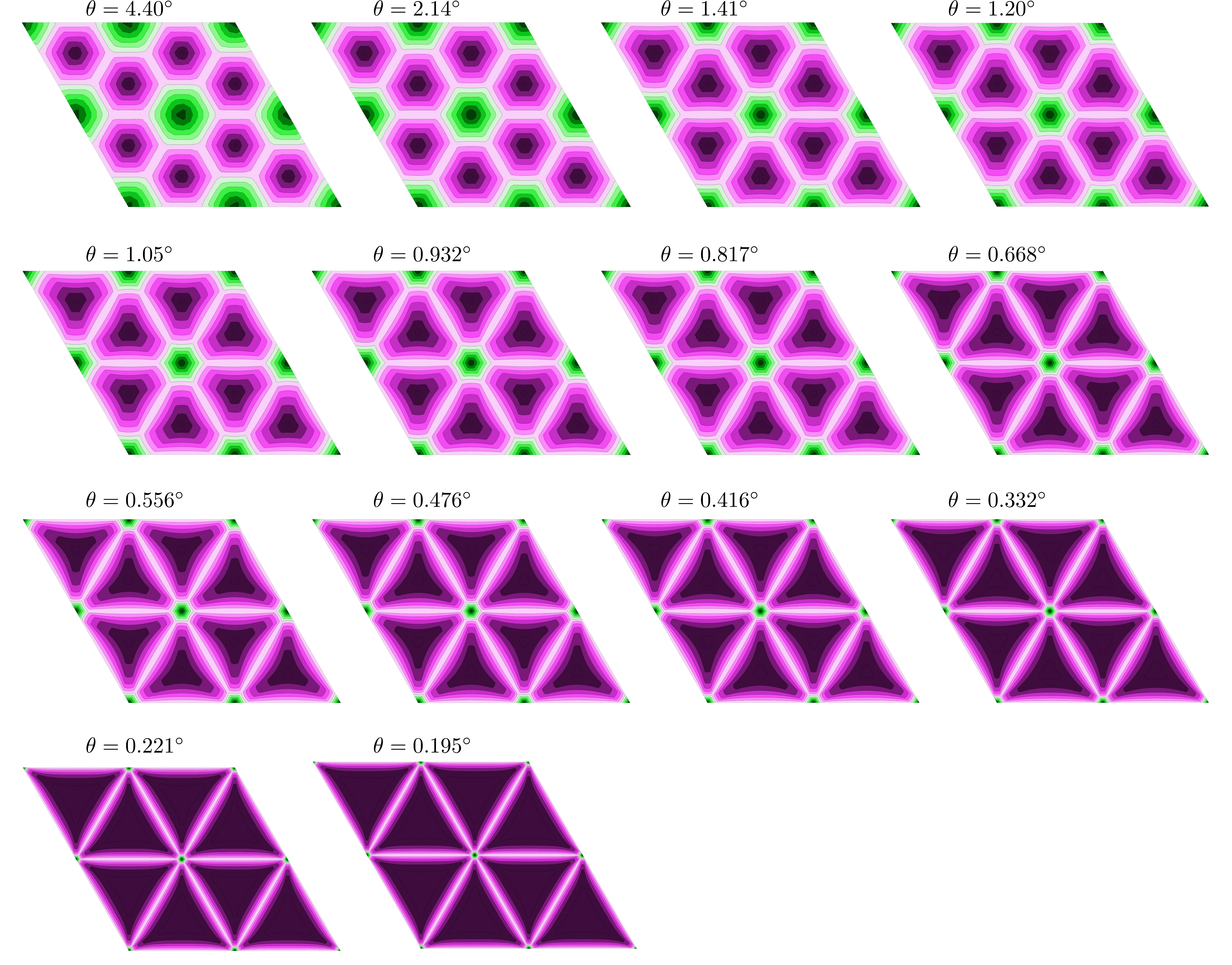}
\caption{A complete representation of the alignment measure for a flat relaxed bilayer, defined as in our previous work \cite{GuineaWalet2019}. The green areas are $AA$ aligned; the purple ones $AB$. White means equal alignment. This is not to scale: the physical dimensions of these figures increase as we go to smaller angles, and the size of the green areas is roughly similar in all cases.}\label{fig:relaxed2}
\end{figure*}

\begin{figure*}
\includegraphics[width=0.99\textwidth ]{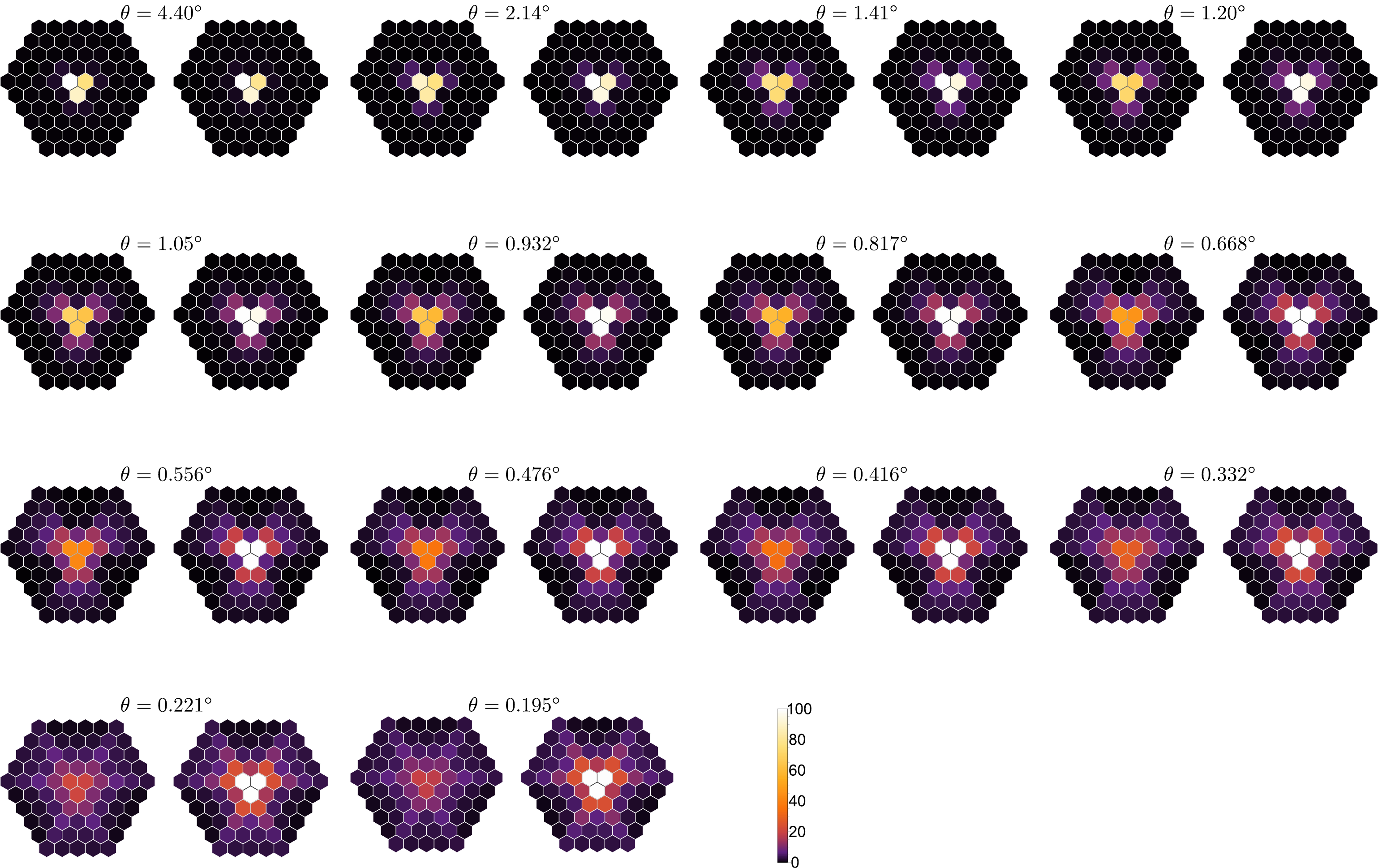}
\caption{The magnitude of the projected couplings for the continuum model for a flat relaxed bilayer: $AA$ couplings on the left and $AB$ to the right.}\label{fig:hex2}
\end{figure*}

\begin{figure*}
\includegraphics[width=0.95 \textwidth]{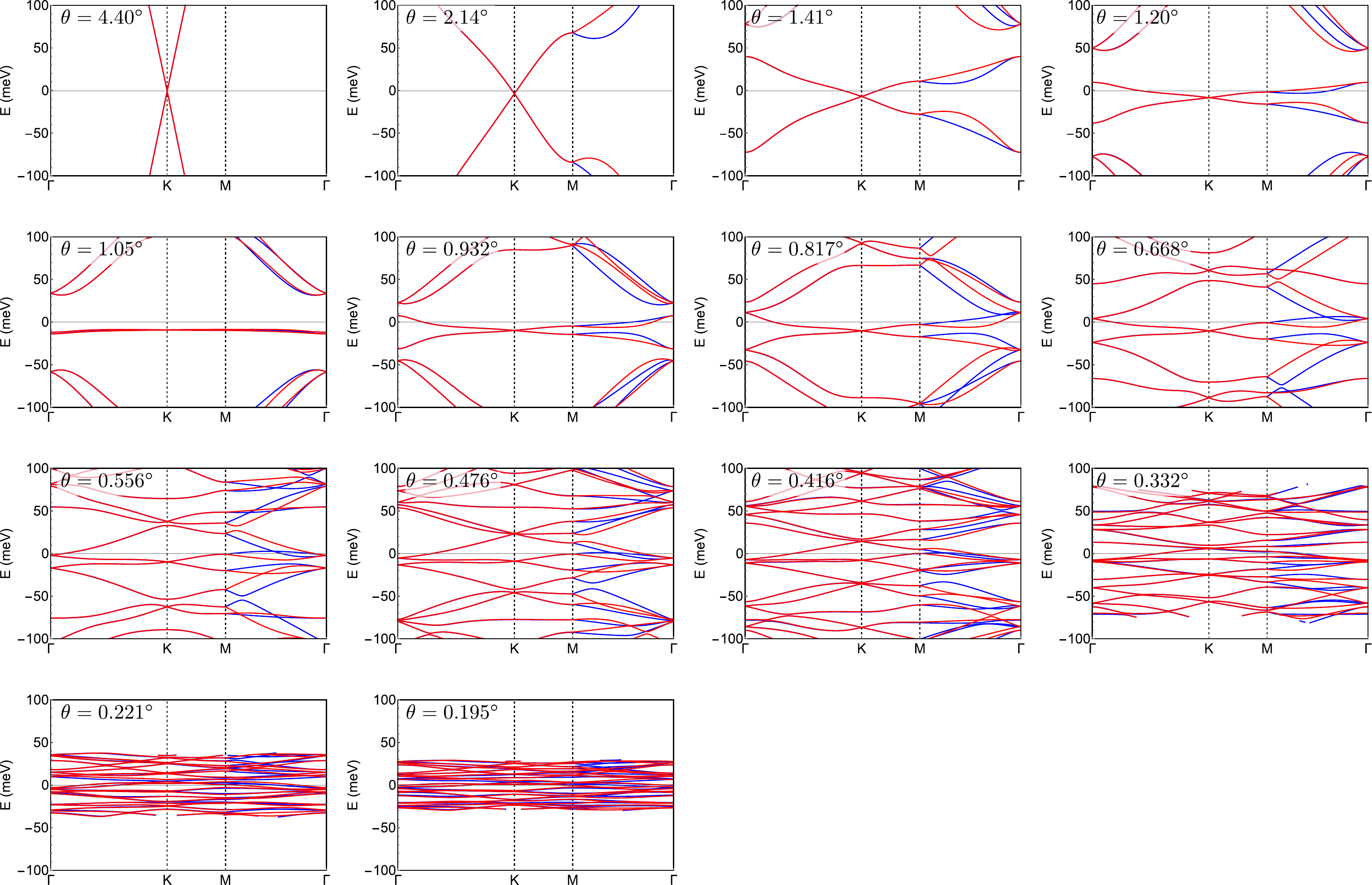}
\caption{Bands for the continuum projected models for a flat relaxed bilayer. Red and blue label the two valleys.}\label{fig:band1}
\end{figure*}
\begin{figure*}
\includegraphics[width=\textwidth]{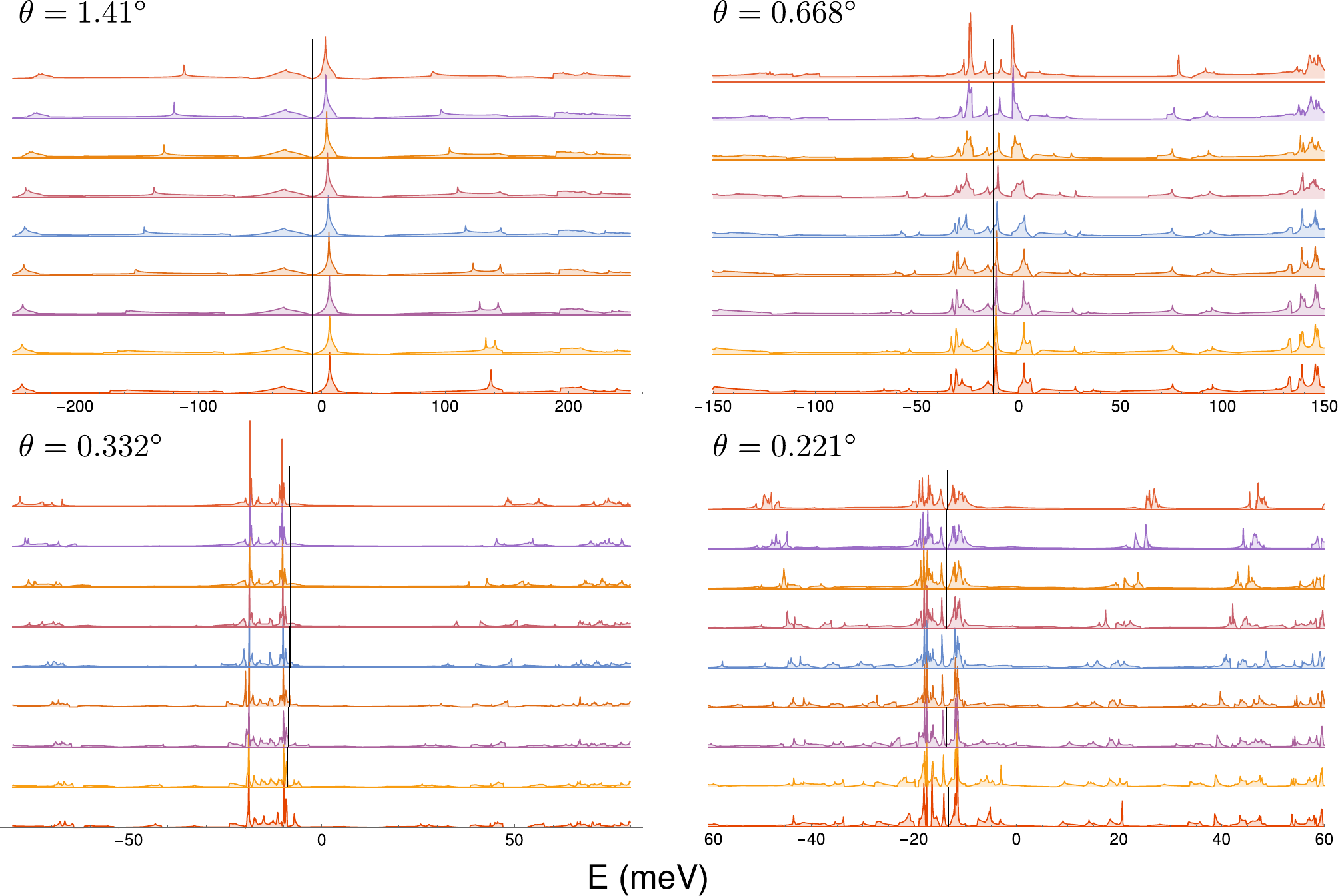}
\caption{Density of states for the continuum projected models with a bias ranging $0$ to $200\,\text{meV}$ for flat undeformed layers. The black lines indicate the Fermi level }\label{fig:DOSbias1a}
\end{figure*}

\begin{figure*}
\includegraphics[width=\textwidth]{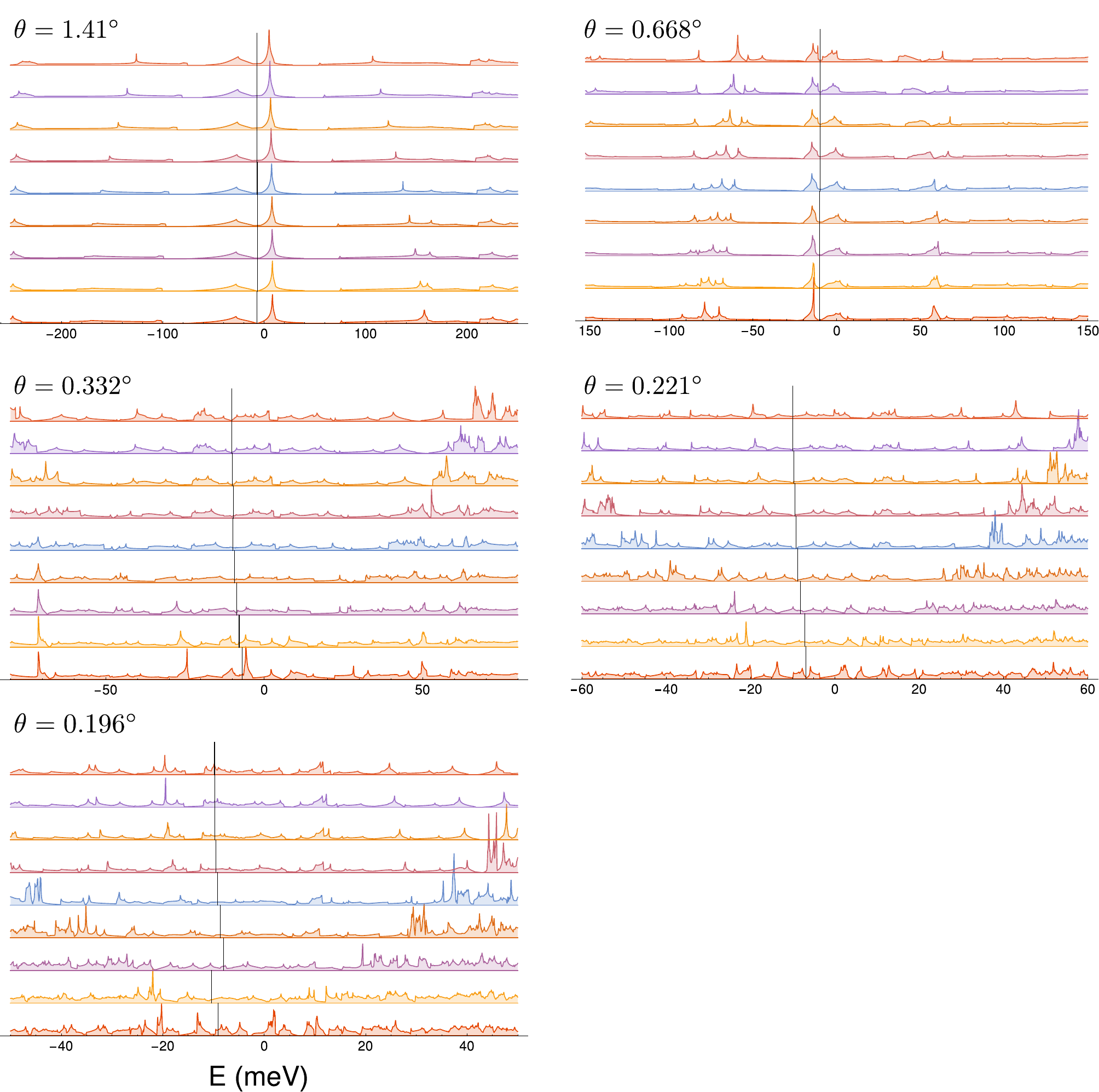}
\caption{Density of states for the continuum projected models with a bias ranging $0$ to $200\,\text{meV}$ for a flat relaxed bilayer.  The black lines indicate the Fermi level}\label{fig:DOSbias2a}
\end{figure*}

\begin{figure*}
\centerline{\href{https://theory.physics.manchester.ac.uk/~mccsnrw/DOS_movie.mp4}{\includegraphics[width=0.7\textwidth]{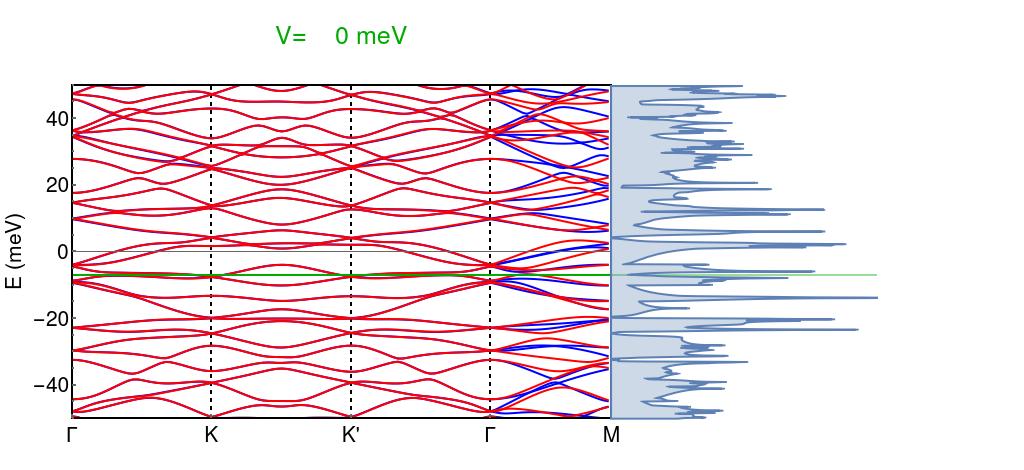}}}
\caption{\label{vid:interest}Video of the development of the edge states for increasing bias; on the left red and blue lines label the states in either valley [the $K$ and $K'$ labels refer to the one valley only (blue lines); they should be mirrored for the other valley], on the right we have the density of states [\url{https://theory.physics.manchester.ac.uk/~mccsnrw/DOS_movie.mp4}]}
\end{figure*}

\begin{figure*}
\includegraphics[width=0.90 \textwidth]{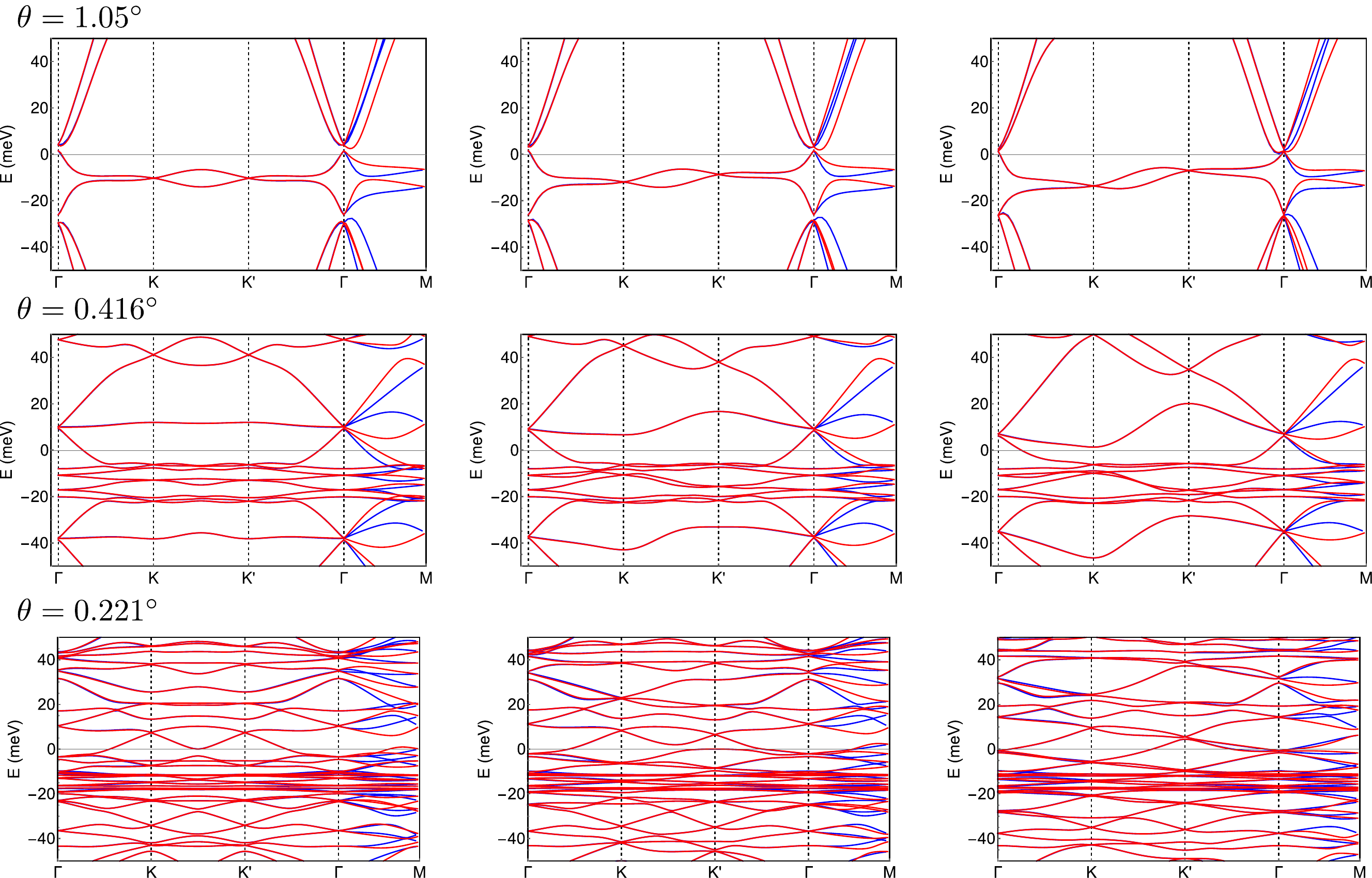}
\caption{Bands for the continuum projected models with a bias ($0$, $50$ and $100\,\text{meV}$) for flat undeformed layers. We have chosen the contour $\Gamma$-$K$-$K'$-$\Gamma$-$M$, which shows the symmetry breaking between the two $K$ points when a bias is present. Red and blue label the two valleys. For simplicity we have labeled point with the same spectrum in both valleys as either $K$ or $K'$}\label{fig:bias1a}
\end{figure*}

\begin{figure*}
\includegraphics[width=0.94\textwidth]{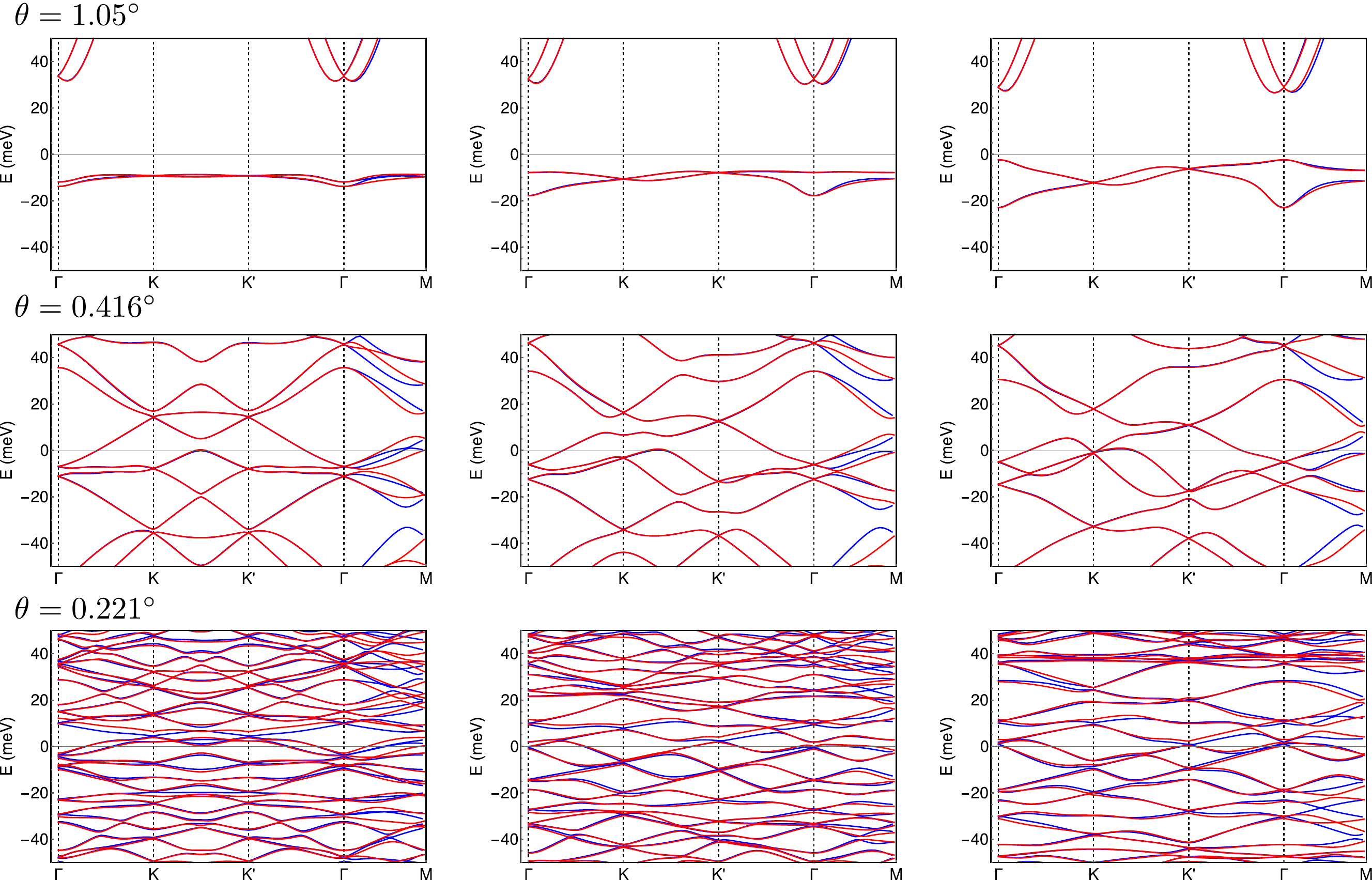}
\caption{Bands for the continuum projected models with a bias ($0$,  $50$, and $100\,\text{meV}$) for a flat relaxed bilayer. Red and blue label the two valleys.}\label{fig:bias2a}
\end{figure*}

\begin{figure*}
\includegraphics[width=\textwidth]{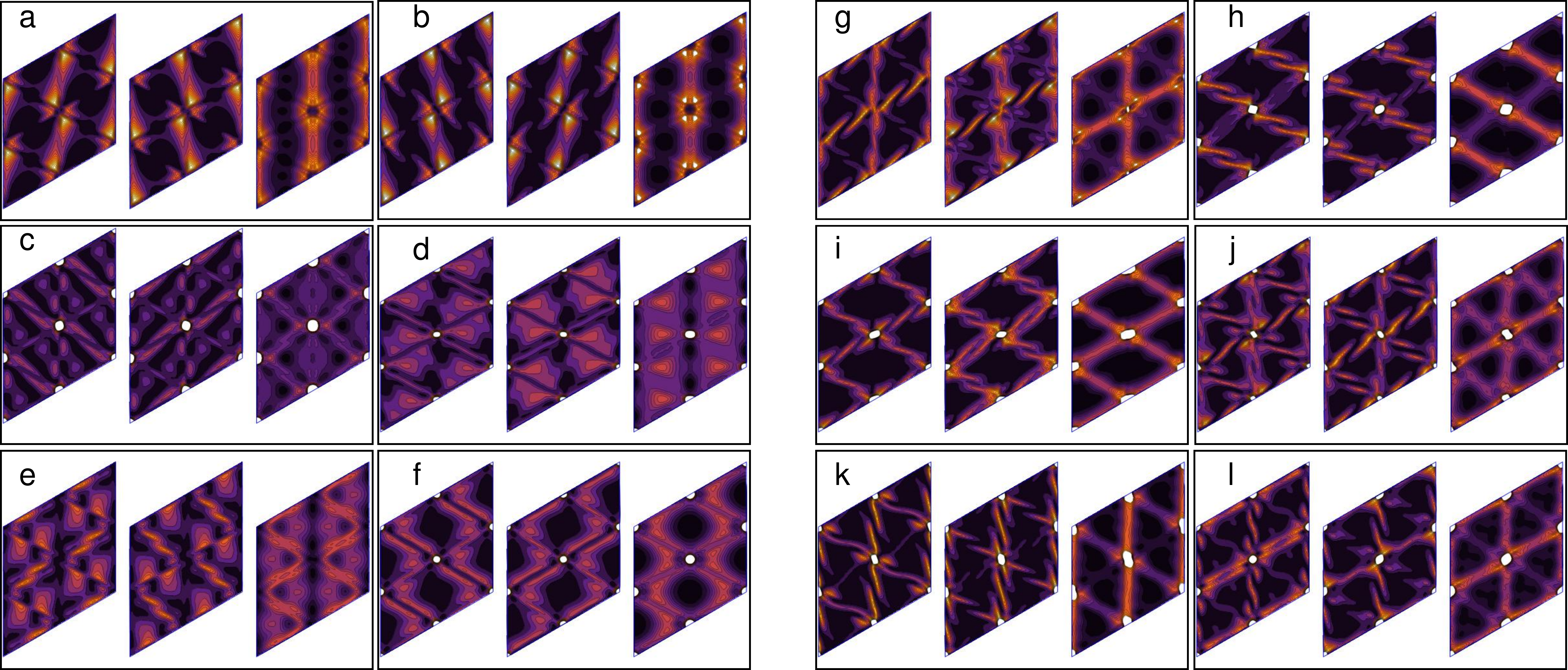}
\caption{The probability density for states near zero energy at the $M$ point for  an angle of $0.221^\circ$. a-f:  \emph{no} applied bias; g-l: bias of $100\,\text{meV}$.
The energy eigenvalues are a) $E=-10.2$ b) $E=-3.95$ c) $E=0.90$ d) $E=2.38$ e) $E=6.36$ and f) $E=8.14\,\text{meV}$; g) $E=-16.09$ h) $E=-14.81$ i) $E=-4.79$ j) $E=-3.97$ k) $E=3.49$ and l) $E=11.75\,\text{meV}$. In each panel left is the top-layer wave function, middle the bottom layer one, and the right panel is the total density.\label{fig:palla}}
\end{figure*}

\end{widetext}
\end{document}